\documentstyle[12pt]{article}
\newcommand{\ket}[1]{{\vert #1\rangle}}
\newcommand{\nn}{\nonumber}
\newcommand{\be}{\begin{equation}}
\newcommand{\ee}{\end{equation}}
\newcommand{\ba}{\begin{eqnarray}}
\newcommand{\ea}{\end{eqnarray}}
\newcommand{\half}{{1 \over 2}}
\begin{document}
\begin{titlepage}

\begin{flushright}
UU-ITP 29/93 \\
HU-TFT-93-60 \\
hep-th/9311102
\end{flushright}

\vskip 0.7truecm

\begin{center}
{ \bf ON THE INFRARED LIMIT OF TWO DIMENSIONAL QCD \\  }
\end{center}

\vskip 1.5cm

\begin{center}
{\bf Antti J. Niemi $^{*}$ } \\
\vskip 0.4cm
{\it Department of Theoretical Physics, Uppsala University
\\ P.O. Box 803, S-75108, Uppsala, Sweden }
\vskip 0.4cm
and \\
\vskip 0.4cm
{\bf Pirjo Pasanen $^{**}$ } \\
\vskip 0.4cm
{\it Research Institute for Theoretical Physics \\
P.O. Box 9, FIN-00014 University of Helsinki, Finland}\vskip 0.2cm
\end{center}

\vskip 2.7cm

\rm
\noindent
We study the infrared limit of two dimensional QCD, with massless
dynamical
Dirac fermions that are in the fundamental representation of the gauge
group.
We find that the theory reduces to a spin generalization of the Calogero
model
with an additional magnetic coupling which is of the Pauli type.
\vfill

\begin{flushleft}
\rule{5.1 in}{.007 in}\\
$^{*}$ {\small E-mail: NIEMI@RHEA.TEORFYS.UU.SE \\}
$^{**}$ {\small E-mail: PIRJO.PASANEN@HELSINKI.FI \\ }
\end{flushleft}

\end{titlepage}

\vfill\eject

\baselineskip 0.65cm

Recently two-dimensional Yang-Mills theories have been extensively
discussed,
in particular in connection of string theories \cite{GroTay_QCDStr}. The
inclusion of matter fields have also been considered
\cite{LanSem_GaugeCyl}, in
particular in the large-$N$ limit \cite{BhDeKle,MinPol}. These
investigations
have revealed several interesting connections between two dimensional
Yang-Mills theories and integrable models \cite{GorNek_CaloYM}.

In the present Letter we shall investigate two dimensional QCD with
massless
Dirac fermions in the fundamental representation of the gauge group SU(N)
for
arbitrary N. We are particularly interested in the infrared weak coupling
limit. We shall quantize the fermions in this limit, and by explicitly
constructing the Fock states we then obtain an effective one dimensional
theory. This effective theory is essentially a spin generalization of the
Calogero model, that has been recently investigated in
\cite{BeGaHaPa_YBSpin,Poly}: In addition of the long range Calogero-spin
interaction we also have a  nearest neighbor spin interaction which
resembles
the Pauli-type coupling between spin and external magnetic field.

\vskip 0.5cm

In the hamiltonian approach, the action for the two dimensional QCD with
massless dynamical Dirac fermions is
\be
S ~=~ \int tr\{ E {\dot A}_{1}  - \frac{1}{2} E^{2} + A_{0} ( \partial
\cdot E
+ ig_{2}[A_{1},E])  + \bar\psi \gamma^{\mu} D_{\mu} \psi \}
\label{action}
\ee
The fermions are in the fundamental representation of the gauge group
SU(N).
For the $\gamma$-matrices we use the representation $\gamma^{0} =
\sigma^{1}$,
$\gamma^{1} = i\sigma^{2}$ and $\gamma^{5} = - \gamma^{0} \gamma^{1}$ in
terms
of the Pauli matrices $\sigma^{i}$. By varying the action with respect to
the
Lagrange multiplier $A_{0}$, we find the Gauss law constraint
\be
\partial_{x} E_{ab} + ig_{2} [A_{1}, E]_{ab} + g_{2} (\psi^{\dagger}
\otimes
\psi)_{ba} ~=~ 0
\label{gauss}
\ee
with $E_{ab}$,  $(A_{1})_{ab}$ and $(\psi^{\dagger} \otimes
\psi)_{ab}$ traceless matrices.

We take the space to be a cylinder with circumference $R$. By proper
rescaling
of the fields and coupling constant, we expect that in the $R \to 0$ limit
the
theory reduces to an effective one dimensional theory. For this we remind
that
in a $D$ dimensional space-time the dimensionalities of the fields and the
coupling constant, in terms of a length scale $L$ are
\be
\left.\begin{array}{ll}
[E]  \sim L^{-\frac{D}{2}}, \qquad
	&[A]  \sim L^{\frac{2-D}{2}}  \cr
[\psi ] \sim L^{\frac{1-D}{2}},\quad
	&[g_{D}] \sim L^{\frac{D-4}{2}}
\end{array}\right.
\ee
If we then Fourier-expand the fields as
\ba
A (x,t) ~&=&~ \frac{1}{\sqrt{R}} \sum_{n} {\cal A}_{n}(t) \exp\{ 2\pi i n
\frac{x}{R} \} \nn \\
E(x,t) ~&=&~ \frac{1}{\sqrt{R}} \sum_{n} {\cal E}_{n}(t) \exp\{ 2\pi i n
\frac{x}{R} \} \label{fourier}\\
\psi (x,t) ~&=&~ \frac{1}{\sqrt{R}} \sum_{n} \psi_{n}(t) \exp\{ 2\pi i n
\frac{x}{R} \} \nn
\ea
where ${\cal A}_{n}$ and ${\cal E}_{n}$ are traceless hermitean matrices
and
$\psi_{n}$ Dirac spinors, and define a new coupling constant
\be
g ~\equiv~ g_{1} = { g_{2} \over \sqrt{R} }
\ee
the Fourier components and the redefined coupling constant $g$ then have
the
proper dimensionalities for a one dimensional theory.

\vskip 0.4cm
In terms of the one dimensional variables,  the Gauss law becomes
\ba
i[{\cal A}_{o}, {\cal E}_{o} ] + {\psi_{o}}^{\dagger} \otimes
{\psi_{o}} + \sum_{n} ( i[ {\cal A}_{n}, {\cal E}_{-n} ] +
{\psi_{n}}^{\dagger}
\otimes \psi_{-n} )
{}~&=&~ 0  \nn \\
{ 2\pi i n \over R } {\cal E}_{n} +  g  \left\{
i[{\cal A}_{o} , {\cal E}_{n} ] + i[{\cal A}_{n}, {\cal E}_{o} ] +
{\psi_{o}}^{\dagger} \otimes
\psi_{n} + {\psi_{n}}^{\dagger} \otimes \psi_{o}
\right. +~ && \\
+ \sum_{m\not= n} \left. \left( i[{\cal A}_{m} , {\cal E}_{n-m}] +
{\psi_{m}}^{\dagger} \otimes \psi_{n-m}
\right) \right\}
{}~&=&~ 0
\nn
\ea
and in the double scaling limit
\be
R ~\to~ 0 ~; \qquad\qquad g_{2} \to 0 ~;
\qquad \qquad
{ g_{2} \over \sqrt{R} } ~=~ g \quad {\rm (finite) }
\label{scaling}\ee
it reduces to
\ba
{\cal E}_{n} ~&=&~ 0  \quad (n \ne 0)  \nn \cr
i[ {\cal A}_{o} , {\cal E}_{o} ] + {\psi_{o}}^{\dagger} \otimes \psi_{o} +
\sum_{n} {\psi_{n}}^{\dagger} \otimes \psi_{-n} ~&=&~ 0   \cr
i[ {\cal A}_{n} , {\cal E}_{o} ] + {\psi_{o}}^{\dagger} \otimes \psi_{n} +
{\psi_{n}}^{\dagger} \otimes \psi_{o} + \sum_{m\not= n}
{\psi_{m}}^{\dagger}
\otimes \psi_{n-m} ~&=&~ 0 \nn
\ea
Furthermore, by investigating the corresponding Fourier-mode expansion of
the
Hamiltonian in (\ref{action}) we observe that the non-constant modes
$\psi_{n}$
($n\not= 0)$ receive an effective mass which is {\it inversely}
proportional to
$R$. In the $R\to 0$ limit these modes then decouple from the (massless)
constant modes and the hamiltonian simplifies to
\be
{\cal H} ~ {\longrightarrow} ~ \frac{1}{2} {\cal E}_{o,ab}
{\cal E}_{o,ba} ~+~ g {\cal A}_{o,ab} ({\psi_{o}}^{\dagger} \otimes
\gamma^{5}
\psi_{o})_{ba}
\label{redham}
\ee
and consequently we can reduce the Gauss law further into
\be
i[{\cal A}_{o} , {\cal E}_{o}]_{ab} +  ({\psi_{o}}^{\dagger} \otimes
\psi_{o})_{ba} ~=~ 0
\label{redgauss}
\ee

In order to solve for the Gauss law (\ref{redgauss}), we first separate it
into
its off-diagonal part
\be
 i[{\cal A}_{o}, {\cal E}_{o}]_{ab} + ({\psi_{o}}^{\dagger} \otimes
\psi_{o})_{ba} =0 \quad  (a\ne b)
\label{offGauss}
\ee
and diagonal part
\be
 i[{\cal A}_{o}, {\cal E}_{o}]_{aa} + ({\psi_{o}}^{\dagger} \otimes
\psi_{o})_{aa} =0.
\label{diagGauss}
\ee

We solve for the off-diagonal part by first gauge transforming  the
constant
mode ${\cal A}_{o}$ into a diagonal matrix
\be
{\cal A}_{o} = {\rm diag}(q_1, q_2, \ldots , q_N)
\label{gaugea}
\ee
Since the Gauss law only eliminates small gauge transformations, these
diagonal
elements are only defined modulo large gauge transformations {\it i.e.}
\be
q_{a} ~\sim~ q_{a} \quad {\rm mod\, } ( {2 \pi n_{a} \over g R  } )
\label{integers}
\ee
where $n_{a}$ are integers. Furthermore, since ${\cal A}_{o}$ is traceless
these diagonal elements are subject to the constraint
\be
\sum_{a} q_{a} ~=~0
\label{qconst}
\ee

For the constant mode of the electric field we write
\be
{\cal E}_{o} = {\rm diag}(p_1, p_2, \ldots , p_N) ~+~ L
\ee
where $L_{ab}$ is off-diagonal and the diagonal elements $p_{a}$ satisfy
the
constraint
\be
\sum_{a} p_{a} ~=~0
\label{pconst}
\ee
{}From (\ref{offGauss}) we then find for the off-diagonal elements
\be
L_{ab} = { i (\psi_{o}^\dagger \otimes \psi_{o})_{ba} \over (q_a - q_b) }
\ee
We substitute this into the hamiltonian (\ref{redham}), and summing over
the
integers $n_{a}$ in (\ref{integers}) we get
\[
{\cal H} ~=~ \half \sum_a  p_a^2 ~+~ { g^{2} R^{2} \over 8 } \sum_{a\ne b}
{
(\psi_{o}^\dagger \otimes \psi_{o})_{ba} (\psi_{o}^\dagger \otimes
\psi_{o})_{ab} \over \sin^{2} [
\half g R ( q_a - q_b ) ]  }
{}~+~ g \sum_a q_a (\psi_o^\dagger \otimes \gamma^5 \psi_o )_{aa}
\]

\be
{\buildrel{ R \to 0} \over {\longrightarrow}} \quad
\half \sum _a  p_a^2 ~+~ \half \sum_{a\ne b} \frac{ (\psi_{o}^\dagger
\otimes
\psi_{o})_{ba} (\psi_{o}^\dagger \otimes \psi_{o})_{ab}}
{ ( q_a - q_b )^{2} }
{}~+~ g \sum _a q_a (\psi_o^\dagger \otimes \gamma^5 \psi_o )_{aa}
\label{newham}
\ee
This hamiltonian is subject to the second class constraint (\ref{qconst}),
(\ref{pconst}). Notice that as a consequence of (\ref{integers}), in the
$R \to
0$ limit the periodic variables $q_{a}$ become defined on the entire real
line.

After diagonalizing ${\cal A}_o $, the diagonal part of the Gauss law
(\ref{diagGauss}) reduces to
\be
(\psi_{o}^{\dagger} \otimes \psi_{o})_{aa} ~=~0
\label{gaussdiag}
\ee
In order to solve for this, we  introduce the chiral components of the
Dirac
spinor
\be
\psi_{oa} = \left ( \matrix{u_a\cr v_a \cr }
	\right )
\ee
We then second quantize these components with the nonvanishing
anticommutators
\be
[ {\bf u}_{a}^{\dagger}, {\bf u}_{a} ]_{+} ~=~ [ {\bf v}_{a}^{\dagger} ,
{\bf
v}_{a} ]_{+}
{}~=~ 1
\ee
and define the corresponding Fock states $\ket{ n^{u}_{a} , n^{v}_{a} }$
with
$n^{u}_{a}, n^{v}_{a} = \pm 1$ by
\ba
{\bf u}_{a} \ket{ +_{a} , n^{v}_{a} } ~=~ \ket{ -_{a} , n^{v}_{a} }
, \qquad \quad & {\bf u}_{a}  \ket{ -_{a} , n^{v}_{a} } ~=~ 0 \nn \\
{\bf u}_{a}^{\dagger} \ket{ -_{a}, n^{v}_{a} } ~=~ \ket{ +_{a} , n^{v}_{a}
},
 \qquad \quad  &  {\bf u}_{a}^{\dagger} \ket{ +_{a} , n^{v}_{a} } ~=~ 0
\ea
so that
\be
[{\bf u}_{a}^{\dagger} , {\bf u}_{a} ]
\hskip 0.1cm
\ket{ \pm_{a} ,  n^{v}_{a} }  ~=~ \pm \ket{ \pm_{a} , n^{v}_{a} }
\ee
and similarly for ${\bf v}_{a}$ and ${\bf v}_{a}^{\dagger}$.
We then introduce a representation of the SU(N) Lie algebra such that
elements
$H_{i}$ $(i=1,...,N-1)$ in the Cartan subalgebra become $(H_{i})_{kl} =
\delta_{ik} \delta_{il} - \delta_{i+1,k} \delta_{i+1,l}$. In this
representation the fermion bilinear in (\ref{redgauss}) can be represented
as
\be
{\psi_{o}}^{\dagger} \otimes \psi_{o}
=  \left( \begin{array}{cccc}
\psi_{1}^{\dagger} \psi_{1} - \psi_{2}^{\dagger} \psi_{2} &
\psi_{1}^{\dagger}
\psi_{2} & . & . \\
\psi_{2}^{\dagger} \psi_{1} & - \psi_{1}^{\dagger} \psi_{1} + 2
\psi_{2}^{\dagger} \psi_{2} - \psi_{3}^{\dagger} \psi_{3} & . & . \\
. & . & . & . \\
. & . & . & - \psi_{N-1}^{\dagger} \psi_{N-1} + \psi_{N}^{\dagger}
\psi_{N}
\end{array} \right)
\label{bilin}
\ee
In particular, for second quantized fermions the diagonal elements in
(\ref{bilin}) can be re-arranged in the normal ordered form
\be
(\psi_{o}^{\dagger} \otimes \psi_{o})_{aa} ~=~
\frac{1}{2} [ \psi_{oa}^{\dagger} ,\psi_{oa} ]
\ee
without additional $c$-number terms. For the physical states the condition
\be
\frac{1}{2} [ \psi_{oa}^{\dagger} , \psi_{oa} ] \ket{ {\rm physical} }
 ~=~ 0
\ee
then reduces to
\ba
n^{u}_{1} + n^{v}_{1} & = & n^{u}_{2} + n^{v}_{2} \nn \\
2 ( n^{u}_{a} + n^{v}_{a} ) - ( n^{u}_{a-1} + n^{v}_{a-1} ) - (
n^{u}_{a+1} +
n^{v}_{a+1} ) & = & 0 \quad (a = 2,...,N-1)
\nn \\
n^{u}_{N-1} + n^{v}_{N-1} & = & n^{u}_{N} + n^{v}_{N}
\ea
The solutions to these equations fall into two different categories:
In the first category which we call an alternating phase, there are
$2^{N}$
states defined by the condition
\be
n^{u}_{a} + n^{v}_{a} ~=~ 0
\ee
for each $a$. In the second category which we call an ordered phase, there
are
only two states. One of them is defined by
\be
n^{u}_{a} ~=~ n^{v}_{a} ~=~ +1
\label{pos}
\ee
and the other by
\be
n^{u}_{a} ~=~ n^{v}_{a} ~=~ -1
\label{neg}
\ee
for all $a$. In the alternating phase, for each $a$ the states in the
fermionic
Hilbert-space are either of the form $\ket{ + ,-}_{a}$
or of the form $\ket{-,+}_{a}$, {\it i.e.} there is an equal number of
$\pm$
configurations. In the ordered phase we have only either $\ket{+,+}_{a}$
or
$\ket{-,-}_{a}$ ({\it i.e.} $+$ or $-$) configurations. Consequently
alternating states differ from ordered states by a rearrangement of N
fermions,
and in the large-N limit the  Hilbert-space decouples into three
subspaces, one
describing the $2^{N}$ dimensional alternating phase and the two others
describing the one dimensional (in terms of the fermion degrees of
freedom)
ordered phases $+$ and $-$. For this reason, in the following we shall
consider
the restrictions of the hamiltonian either to the alternating or to the
ordered
subspaces, also for finite values of N.

In the ordered phases the dynamics is trivial: the hamiltonian
(\ref{redham})
simply reduces to the free hamiltonian
\be
{\cal H} ~\to~ \half \sum_{a} p_{a}^{2}
\ee

In the alternating phase, the physical states are linear combinations of
the
form
\be
\ket { {\rm physical} } ~=~ \sum\limits_{n_{a} = \pm}
\Phi_{n_{1} ... n_{N}} (q) \bigotimes_{a=1}^N
\ket{ n_{a} , - n_{a} }
\label{hilbert}
\ee
In this Hilbert-space we can further simplify the Hamiltonian
(\ref{newham}).
For this, we introduce the following operators
\ba
H_a   &=& \half ( u^{\dagger}_a u_a - v^{\dagger}_a v_a )\cr
E^+_a &=& u^\dagger_a v _a\cr
E^-_a &=& v^\dagger_a u _a\cr
C_a   &=&  \half (u^\dagger_a u _a+ v^\dagger_a v_a )
\label{sl2rep}
\ea
These operators act only on the $a^{th}$ fermionic states and satisfy the
SU(2)
Lie-algebra commutation relations
\be
[H_a, E^{\pm}_b] = \pm\delta_{ab}E^{\pm}_a  \qquad \qquad
[E^+_a, E^-_b] = 2\delta_{ab} H_a .
\label{sl2}
\ee
and the Casimir operators $C_a$ are proportional to the unit operator,
\be
C_{a} ~=~ \frac{1}{2} \cdot I_{a}
\ee
In particular, we can identify (\ref{sl2})  as the {\it
fundamental} representation of SU(2). For generators in the {\it
fundamental}
representation the  permutation operator $P_{ab}$ for spins at sites $a$
and
$b$ can be expressed as
\be
P_{ab} = \half I_a I_b + 2 H_a H_b + E^+_a E^-_b + E^-_a E^+_b
\label{permutation}
\ee
and if we define normal ordering in the second term of (\ref{newham}) by
\be
: (\psi_{o}^\dagger \otimes \psi_{o})_{ba} (\psi_{o}^\dagger \otimes
\psi_{o})_{ab} : ~=~
(  {\bf u}^{\dagger} {\bf u} + {\bf v}^{\dagger} {\bf v} )_{ba}
(  {\bf u}^{\dagger} {\bf u} + {\bf v}^{\dagger} {\bf v} )_{ab}
 \qquad (a\ne b)
\ee
we can rewrite this second term as
\ba
{\cal H}_{2} &~=~&  \half\sum_{a\ne b}^{N}{ 1 \over (q_{a} - q_{b})^{2} }
\{  C_b I_a + C_{a} I_{b}- 2 C_b C_a - 2 H_aH_b - E^+_b E^-_a - E^-_bE^+_a
\}
\nn \\
&=~& \half\sum_{a\ne b}^{N}{ \frac{(1- P_{ab})}{(q_a - q_b)^2}}
\ea

Similarly, by using the representation (\ref{sl2rep}) we can also simplify
the
third term in the hamiltonian (\ref{newham}) into
\be
{\cal H}_{3} ~=~ g \sum_{a} q_{a} ({\psi_{o}}^{\dagger} \otimes
\gamma^{5} \psi_{o})_{aa}  ~=~
2g\sum_{a=1}^{N-1}{ (q_a -q_{a+1})( H_a - H_{a+1})}
\ee
Our final hamiltonian in the alternating phase is then
\be
{\cal H} =\half\sum_a^N p_a^2 ~+~ \half\sum_{a\ne b}^{N}
{ 1- P_{ab} \over (q_a - q_b)^2 } ~
+~ 2g\sum_{a=1}^{N-1}{ (q_a -q_{a+1})( H_a - H_{a+1})}
\label{final}
\ee
In the $g\to 0$ limit this hamiltonian reduces to the
integrable long range spin interaction hamiltonian  discussed in
\cite{BeGaHaPa_YBSpin,Poly}. We have tried to show that our hamiltonian is
also
integrable by using the methods developed
\cite{BeGaHaPa_YBSpin,Poly,BrHaVas}.
However, due to complications that are caused by the {\it linear}
Pauli-type
term we have not yet succeeded in establishing whether (\ref{final})
indeed is
integrable. There are nevertheless a number of interesting, simple
properties
that we wish to record:

The following two spin states can be viewed as ground state
spin
configurations of (\ref{final}),
\ba
\ket{\rm vac }_{+-} ~=~ ... \ket{ +- } \ket{+- } \ket{+-} ...
\nn \\
\ket{ \rm vac }_{-+} ~=~ ... \ket{ -+ } \ket{-+ }\ket{-+} ...
\label{vacua}
\ea
In particular, the second and third terms in (\ref{final}) both vanish on
these
states. Consequently the ground state of (\ref{final})
appears degenerate, in analogy with {\it e.g.} the degeneracy of the Ising
model ground state in the magnetic phase. For finite N we have finite
energy
domain wall configurations which interpolate between these two vacuum
states,
for example
\be
  ... \ket{+-} \ket{+-} \ket{+-} \ket{-+} \ket{-+}
\ket{-+} ...
\ee
is a domain wall configuration which interpolates from $\ket{\rm
vac}_{+-}$ to
$\ket{\rm vac}_{-+}$.  As N$\to \infty$ the energy of these domain walls
diverges, and we obtain two disjoint Hilbert-spaces.  To see this, we
estimate
the domain wall energy
as follows: By translation invariance, it is sufficient to consider a
linear
chain with the spin $q_{a}$ at the position $n_{a} \cdot \Delta$ ($n_{a} =
0,
\pm 1, \pm2, ...$) with a domain wall located at the origin. The energy of
the
domain wall can then be estimated from the second term in (\ref{final}),
by
observing that due to the spin permutation operator there is a
contribution
only from terms that connect different sides of the origin. By direct
substitution we find that the energy of the domain wall diverges for $N\to
\infty$,
\be
E ~\sim~ \sum\limits_{q_{a} q_{b} < 0} { 1 \over (q_{a} - q_{b})^{2} }
{}~\sim~
\frac{1}{\Delta^{2}} \sum_{ n,m =1 }^{N} {1 \over (n +
m )^{2}} ~\sim~ \ln N ~ {\buildrel {N\to\infty}
\over {\longrightarrow}} ~ \infty
\ee

The hamiltonian (\ref{final}) also admits a bosonized form. For this, we
first
represent the SU(2) generators classically: The coadjoint orbit of SU(2)
coincides with the Riemann sphere $S^{2}$. With coordinates $z$ and $\bar
z$
and Poisson bracket
\be
\{ z, \bar z \} ~=~ \frac{1}{2} (1 + |z|^{2})^{2}
\label{twoform}
\ee
the SU(2) generators are
\ba
\sigma_{3} ~&=&~ - \frac{1}{2}\cdot { 1 - |z|^{2} \over 1 + |z|^{2} }
\nn
\\
\sigma_{+} ~&=&~ { \bar z \over 1 + |z|^{2} }
\label{bosonize}\\
\sigma_{-} ~&=&~ {  z \over 1 + |z|^{2} }
\nn
\ea
and the bosonized version of the action is obtained by simply replacing
for
each $a$ the SU(2) generators in (\ref{final}) by (\ref{bosonize}), and
then
quantizing each canonical pair  $z_{a}$, ${\bar z}_{a}$ using the
symplectic
structure (\ref{twoform}). Such a bosonized realization of (\ref{final})
could
be valuable in the investigation of its integrability. It could also be
useful
more generally, in the investigation of bosonized two dimensional QCD.

\vskip 0.5cm

\vskip 0.5cm
In conclusion, we have investigated two dimensional QCD in the infrared
weak
coupling limit (\ref{scaling}). We have found that with Dirac fermions in
the
fundamental representation of SU(N) we obtain an integrable spin
generalization
of the Calogero model with an additional Pauli type magnetic interaction
term.
We do expect that our model is integrable, however we have not yet been
able to
establish this.

We have also discussed some simple properties of our model. In particular,
we
have found some qualitative differences between the finite N and
N$\to\infty$
cases: For N$\to\infty$ the ordered states (\ref{pos}), (\ref{neg})
decouple
from the spectrum, and in addition the domain walls connecting the two
different vacuum configurations (\ref{vacua}) receive an infinite energy.
In
future publications we hope to report on the consequences of these
effects.

\vskip 1.0cm

We are indebted to L. Faddeev and A. Polychronakos for very useful discussions.
We also wish to thank N. Nekrasov and K. Palo for their helpful comments.


\begin{thebibliography}{9}

\bibitem{GroTay_QCDStr} D.~J.~Gross, W.~Taylor.
{\it  hep-th }9311072 and
{\it Nucl.Phys.}{\bf B400}(1993)181 (hep-th 9301068).

\bibitem{LanSem_GaugeCyl} E.~Langmann, G.~W.~Semenoff. {\it
Phys.Lett.}{\bf
B296} (1992)117 (hep-th  9210011).

\bibitem{BhDeKle} G.~Bhanot, K.~Demeterfi, I.~R.~Klebanov. hep-th 9307111.

\bibitem{MinPol} J.~A.~Minahan, A.~P.~Polychronakos. {\it Phys.Lett.}{\bf
B312}(1993) 155 (hep-th 9303153) and  hep-th 9309044.

\bibitem{GorNek_CaloYM} A.~Gorsky, N.~Nekrasov. hep-th 9304047.

\bibitem{BeGaHaPa_YBSpin} D.~Bernard, M.~Gaudin, F.D.M.~Haldane,
V.~Pasquier.
hep-th 9301084.

\bibitem{Poly} A.~P.~Polychronakos. {\it Phys. Rev. Lett.} {\bf 69} (1992)
703.

\bibitem{BrHaVas} L.~Brink, T.~H.~Hansson, M.~Vasiliev. {\it Phys. Lett.}
{\bf
B286} (1992) 109.





\end{thebibliography}
\end{document}